\documentclass[12pt]{article}

\usepackage{amsfonts,amssymb,amsmath}
\usepackage{graphicx}
\usepackage{url}
\usepackage{tikz}

\usetikzlibrary{arrows,shapes,shadows,fit,backgrounds}
\tikzstyle{trader} = [circle, draw, top color=white, bottom color=blue!30, draw=blue!50!black!100, drop shadow, minimum height=4em]
\tikzstyle{bank} = [rectangle, draw, top color=white, bottom color=red!20, draw=red!50!black!100, drop shadow, rounded corners, minimum height=3em, text width=4em, text centered]
\tikzstyle{market} = [rectangle, draw, top color=white, bottom color=green!20, draw=green!50!black!100, drop shadow, rounded corners, minimum height=3em, text width=4em, text centered]
\tikzstyle{background} = [rectangle,fill=gray!10, inner sep=0.2cm, rounded corners=5mm]
\tikzstyle{line} = [draw, latex'-latex']
\tikzstyle{from} = [draw, latex'-]
\tikzstyle{to} = [draw, -latex']

\usepackage[]{hyperref}
 \hypersetup{
 colorlinks=true, 
 breaklinks=true, 
 urlcolor= blue, 
 linkcolor= blue, 
 bookmarksopen=true, 
 pdfauthor={Pallavicini, A., Brigo D.}
 }

\begin{document}

\title{\vspace{-2cm} 
{\bf \large CCPs, Central Clearing, CSA, Credit Collateral and Funding Costs Valuation FAQ:} \\ {\normalsize Re-hypothecation, CVA, Closeout, Netting,  WWR, Gap-Risk,  Initial and Variation Margins, Multiple Discount Curves, FVA?}\thanks{{\small This dialogue, available at damianobrigo.it, SSRN.com and arXiv.org, is second in a series started with ``Counterparty Risk, Collateral and Funding FAQ", see {\tt{http://ssrn.com/abstract=1955204}}
or {\tt{http://arxiv.org/abs/1111.1331}}}. See also the Wiley book ``Counterparty Credit Risk, Collateral and Funding" \cite{BMP2013}}
}
\author{
Damiano Brigo \\ {\normalsize Dept. of Mathematics }\\ {\normalsize Imperial College London} \\ {\normalsize and Capco Institute}  \and Andrea Pallavicini \\ {\normalsize Dept. of Mathematics }\\ {\normalsize Imperial College London} \\ {\normalsize and Banca IMI} 
}

%
%

\date{\small First Version: Nov 15, 2013. This Version: \today}

\maketitle

\vspace{-1cm}

\begin{abstract}
We present a dialogue on Funding Costs and Counterparty Credit Risk modeling, inclusive of collateral, wrong way risk, gap risk and possible Central Clearing implementation through CCPs. This framework is important following the fact that derivatives valuation and risk analysis has moved from exotic derivatives managed on simple single asset classes to simple derivatives embedding the new or previously neglected types of complex and interconnected nonlinear risks  we address here. This dialogue is the continuation of the ``Counterparty Risk, Collateral and Funding FAQ" by Brigo (2011). In this dialogue we focus more on funding costs for the hedging strategy of a portfolio of trades, on the non-linearities emerging from assuming borrowing and lending rates to be different, on the resulting aggregation-dependent valuation process and its operational challenges, on the implications of the onset of central clearing, on the macro and micro effects on valuation and risk of the onset of CCPs, on initial and variation margins impact on valuation, and on multiple discount curves. Through questions and answers (Q\&A) between a senior expert and a junior colleague, and by referring to the growing body of literature on the subject, we present a unified view of valuation (and risk) that takes all such aspects into account.  
\end{abstract}

\newpage

\pagestyle{myheadings}
\markboth{}{{\footnotesize D. Brigo, A. Pallavicini: CCPs, Funding Costs, Collateral and Credit  FAQ / Q\&A}}

\bigskip

\tableofcontents

\vspace{1cm}

\noindent 
{\bf AMS classification codes: 60G51, 60G70, 60H35, 62G32, 65C05, 65C20, 91B70}

\newpage


\newpage

\begin{center}
\includegraphics{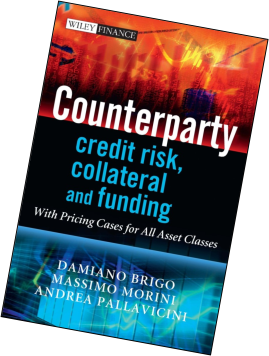}
\end{center}

\bigskip

\noindent {\large See also the first chapter of the Wiley Finance (March 2013) Book
 {\bf ``Counterparty Credit Risk,
Collateral and Funding} with Pricing Cases for all Asset Classes"

\bigskip

\noindent {\large by Damiano Brigo\footnote{This summary FAQ, together with the previous part \cite{Brigo2011faq}, is based on the authors last 8 years of research and more specifically of papers 
\cite{Brigo2005}, \cite{Brigo2006}, \cite{alfonsi}, 
\cite{BrigoChourBakkar}, 
\cite{BrigoBuescuMorini},
\cite{brigoliu2012}, 
\cite{BrigoCapponi}, 
\cite{BrigoCapponiPalla}, 
\cite{brigoetalcollateral}, 
\cite{Brigo08},
\cite{BrigoCousot},
\cite{elbachir}, 
\cite{BrigoMas}, 
\cite{BrMer}, 
\cite{BrigoMorini2006}, 
\cite{BrigoMorini2010}, 
\cite{BrigoMorini2010Flux}, 
\cite{BrigoMorini2011Risk}, 
\cite{BMP2013}, 
\cite{BrigoMoriniTarenghi}, 
\cite{BrigoPalla07}, 
\cite{BrigoPalla08}, 
\cite{BrigoPalla13}, 
\cite{BrigoPallaPapa}, 
\cite{brigopallatorre}, 
\cite{BrigoTarenghi2004}, 
\cite{BrigoTarenghi2005}, 
\cite{Moreni2010},
\cite{Moreni2012},
\cite{BrigoPallaPerini},
\cite{BrigoPallaPerini2}, 
\cite{TorreJRMFI},
supported by several co-authors, including Claudio Albanese, Imane Bakkar, Tom Bielecki, Cristin Buescu, Agostino Capponi, Kyriakos Chourdakis, Nicola Moreni, Massimo Morini, Frank Oertel, Vasileios Papatheodorou, Frederic Patras, Daniele Perini, Mirela Predescu, Marco Tarenghi and Roberto Torresetti. We are grateful to all participants in the dinner at the Royal Society in London on Oct 27, 2011, and to all participants of the Quant Congress Europe 2011, Global Derivatives 2012 and 2013, Risk Minds US 2012, Risk Minds 2013, the Newton Institute meeting in Cambridge on March 28 2013,  and CVA Collateral Damages 2012 in Paris for helpful discussions on CVA and Funding after training courses, plenary talks and over conference dinners. This paper expresses the authors opinion and is in no way representative of the institutions the authors work for or are associated with.}, Massimo Morini and Andrea Pallavicini } 
}



\newpage

\section{A Dialogue on CCPs, Funding, Collateral and Credit Risk} 

Although research on counterparty risk pricing and its extensions, at the time without much collateral or funding costs analysis, started way back in the nineties, with us joining the effort back in 2002, the different aspects of counterparty credit risk and funding costs exploded after the beginning of the financial crisis in 2007 and especially 2008. In less than four years we have seen the emergence of a number of features that the market operators are struggling to account for with consistency. Further, the several possible definitions and methodologies for counterparty risk and funding costs may create confusion. This dialogue is meant to provide a colloquial guide to the different aspects of counterparty risk, collateral modeling and funding costs, especially now that central clearing through CCPs and standard CSA guided by ISDA are being implemented massively. It is in the form of a Q\&A between a CVA expert and a newly hired colleague, and provides detailed references for investigating the different areas sketched here more in detail.

\medskip

{\emph{Important: This dialogue continues ideally from the dialogue in \cite{Brigo2011faq} or in the first chapter of \cite{BMP2013}. We advise the reader who is not familiar with the previous dialogue to have a look at that dialogue, although the present dialogue may be read independently. }}

\section{Funding Costs Pricing: The full replication approach}

\begin{itemize}

\item[Q:] [Junior colleague] We haven't spoken for a while, it must have been one year at least, how have you been?

\item[A:] Quite busy with the issues on funding costs actually, but it would be good to catch up, indeed, I heard you have been doing well in your job in this first year. 

\item[Q:] Not too bad, thanks to your help too. Would you be ok to have a second conversation? I found our first conversation to be quite helpful.

\item[A:] Yes we could have an update on funding probably, and then move on discussing CCPs a little? Would that be of interest? 

\item[Q:] That would be fantastic. Let's book a room here for tomorrow 3pm?

\item[A:] Sounds good!

\item[Q:] [The day after, at 3pm] Ok, we covered quite a lot of stuff in our previous conversation. The inclusion of funding costs is still debated, however, and we might restart from there? 

\item[A:] Indeed. Let's see what we have in the literature. The working paper \cite{Crepey2011}, then published in \cite{Crepey2012a} and \cite{Crepey2012b}, remains one of the most comprehensive theoretical treatments I have seen so far. As we discussed last time, one limitation is that it does not allow for underlying credit instruments in the portfolio, and has possible issues with FX. It is a very technical paper. A related framework that is more general and includes most recent literature as a special case is in \cite{BrigoPallaPerini}, that since our last chat has been expanded and clarified further in \cite{BrigoPallaPerini2}. I think this is the most complete work so far, and it lead also to developments on including funding effects in multiple discount curves, see \cite{BrigoPalla13}, and also to a rigorous pricing of a deal cleared with CCPs, by including the margining fees and residual risks, see \cite{CCPPallaBr2013}.

\item[Q:] Hey slow down please... this is too much in a single sentence and we need to loosen this a little. Let's take a couple of steps back to make sure we followed the same reasoning in this year? 

\item[A:] Good point, let's check if we are aligned at least on the starting points. What is the funding costs pricing problem? To put it in a nutshell, when you need to manage a trading portfolio, you may need to obtain cash and / or assets in order to do a number of operations: maintain the hedge of the position, posting collateral on margin calls, paying interest on collateral received, for example.  This is cash / assets you may obtain from your Treasury department or in the market. You may also receive cash or assets as a consequence of being in the position: a coupon, a notional reimbursement, a positive mark to market move, getting some collateral, a closeout payment, receiving interest on collateral you posted, a good movement in your hedge. All such flows need to be remunerated: if you are borrowing, this will have a cost, and if you are lending, this will provide you with some revenues. Including the cost of funding into your valuation framework means to properly account for such features.

\item[Q:] You mention ``hedging" and ``valuation", but valuation in the replication approach is based on a hedging portfolio that satisfies a self financing condition, at least theoretically. This means you don't need to inject external funds to rebalance the replicating hedge. The hedge can be fully financed thanks only to price movements of cash and risky assets. Now you are telling me that the hedge may be expensive and one may need to fund the hedge explicitly with external funds. Aren't we violating the self financing condition, and if so, are we authorized to use a self financing and pricing approach?

\item[A:] Excellent point, in fact we are not violating the self financing condition but we are extending it to a larger system, that includes our treasury, collateral and market funding policies. There is still conservation, in other terms, but on a larger level, as the idealized risk free rate bank account, namely the numeraire of the risk neutral measure, is no longer a real asset for us.
You may remember the bank account and the risk neutral measure from our previous discussion last year \cite{Brigo2011faq} [see previous dialogue]. 

\item[Q:] But certainly this is stretching the replication approach a lot... can we trade freely into the funding assets and the treasury and collateral accounts? For example, I can understand that with re-hypothecation "on" the collateral account becomes a sort of ``tradeable" resource whose borrowing and lending we can modulate so as to ``trade" with our own treasury department account, but this is not at all straightforward to me. And what happens if you don't have the risk free bank account anymore? How is your hedge implemented?

\item[A:] That's a very good question and it's interesting to list what Pallavicini et al \cite{BrigoPallaPerini} \cite{BrigoPallaPerini2} and Brigo et al \cite{BMP2013} find: even if they start from a risk neutral measure and a risk-free bank account accruing at the risk free rate, once their derivation goes through the risk free rate disappears and only real market funding and collateral rates remain to drive the pricing equations.

\item[Q:] So the risk free rate would be a sort of hidden variable that does not show up in the final valuation formula??

\item[A:] I'd say more an instrumental variable rather than a hidden variable. You know, ''hidden variables" reminds me of the De Broglie-Bohm pilot wave theory in Quantum Mechanics, of Bell's theorem and of the ongoing discussion on the meaning of quantum theory... or another analogy is that of bare mass in quantum field theory, it looks like the risk free rate is like a bare mass but the funding rates are clothed mass...

\item[Q:] Ah, now wouldn't this be a lovely topic to discuss. Why don't we talk about the so called Copenhagen interpretation with its infinite regresses and magic measurements, or many worlds, or advanced action, or...

\item[A:] Don't get carried away! That's another debate entirely. Quantum effects and complex numbers are forbidden in Finance!!! Or it's your job! [grinning]

\item[Q:] It could be useful to have a company that is defaulted and not defaulted at the same time... I don't know, Schroedinger LTD?

\item[A:] You are as funny as a headache on Saturday night, let's get back to the risk free instantaneous interest rate as an instrumental (rather than hidden) variable, instrumental meaning that it does not play any role in the end but is only instrumental for the derivation.

\item[Q:] I don't know... sounds quite strange, if you use it...

\item[A:] Let's look at the derivation. You start by writing the cash flows of your  financial product first, and I mean the basic cash flows associated with the payout and dividends, driven by the key risk factors of your instrument. So if you have an European equity call option the only cash flow is at final maturity and is the payout of the option. If you have a bond you'll have coupons and a final notional reimbursement. Etc.

\item[Q:] Ok so far so good. What next?

\item[A:] You then add the collateral costs, meaning the interest on collateral you need to pay when you receive collateral, or the interest on collateral you receive when you post collateral.

\item[Q:] Makes sense. What else?

\item[A:] You then add the default closeout term including of course the collateral guarantee. Suppose your counterparty defaults first. At default you check if you have enough collateral or not in the account to cover your loss, and you proceed with the collateral netting rules by ISDA. Again, if there is considerable gap risk, you may find a relevant mismatch in a number of scenarios, enough to generate a sizeable residual CVA. The margin period of risk is also key: collateral disputes or other delays in realigning the mark to market through collateral may enlarge the margin period of risk to ten days, for example. In ten days, under a stressed market, the mark to market of a deal can change considerably, and the collateral last posted might become inadequate to cover the loss. 

\item[Q:] Can this effect really be that sizeable? 

\item[A:] We discussed Gap risk already in  our previous conversation one year ago \cite{Brigo2011faq}. However a refresher might indeed be good. Ok, summing up: through margining, the collateral account, at the periodic margining dates, contains an amount that is close to the actual value of the portfolio to the collateral receiver and if one counterparty were to default, the amount would be used by the surviving party as a guarantee (and viceversa).
{\emph{Gap Risk}} is the residual risk that is left, due to the fact that the realignment is only periodical. If the market were to move a lot between two realigning ("margining") dates, a significant loss would still be faced.

\item[Q:] You know, there is this folk belief: Collateral completely kills CVA and gap risk is negligible.

\end{itemize}

\section{Gap Risk}

\begin{itemize}

\item[A:] I am going to show you an example where this is not the case at all. These plots I am showing you in my tablet now are taken from  \cite{brigoetalcollateral} and  \cite{BrigoCapponiPalla}.
First example: Collateralized bilateral CVA for a netted portfolio of interest rate swaps (IRS) with 10y maturity and 1y coupon tenor for different default-time correlations with (and without) collateral re-hypothecation. 

\begin{figure}
\begin{center}
\includegraphics[scale=1]{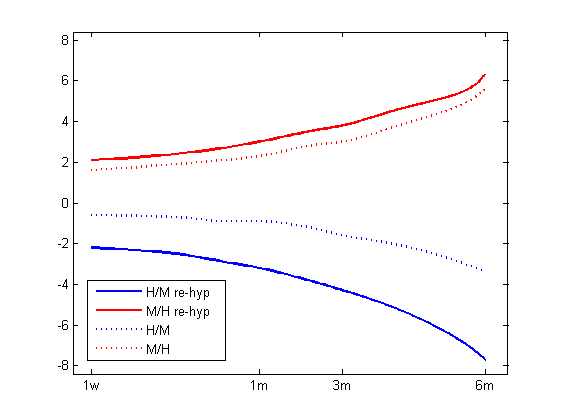} 
\end{center}
\caption{CVA after collateralization (Gap Risk) for a interest rate swap portfolio}\label{fig:cvairs}
\end{figure}

\item[Q:] What does this figure show? 

\item[A:] Figure \ref{fig:cvairs} shows the BVA(namely DVA-CVA) for a ten-year IRS under collateralization through margining as a function of the update frequency $\delta$ with zero correlation between rates and counterparty spread, zero correlation between rates and investor spread, and zero correlation between the counterparty and the investor defaults. The model allows for nonzero correlations as well.
The Continuous lines represent the re-hypothecation case, while {dotted lines} represent the opposite case. The red line represents an investor riskier than the counterparty, while the blue line represents an investor less risky than the counterparty. All values are in basis points.

\item[Q:] I see very small numbers, below 2 basis points? Collateral seems to be very effective here, residual DVA-CVA is almost zero.

\item[A:] Indeed, from the Figure, we see that the case of an investor riskier than the counterparty (M/H) leads to positive value for DVA-CVA, while the case of an investor less risky than the counterparty has the opposite behaviour. If we inspect the DVA and CVA terms as in the paper we see that when the investor is riskier the DVA part of the correction dominates, while when the investor is less risky the counterparty has the opposite behaviour
Re-hypothecation enhances the absolute size of the correction, a reasonable behaviour, since, in such case, each party has a greater risk because of being unsecured on the collateral amount posted to the other party in case of default. But you are right numbers are very small and collateral here is very effective. 

\item[Q:] So folk opinions are right here?

\item[A:] Wait a minute. Let us now look at a case with more contagion: a CDS.

\begin{figure}
\begin{center}
\includegraphics[scale=0.4]{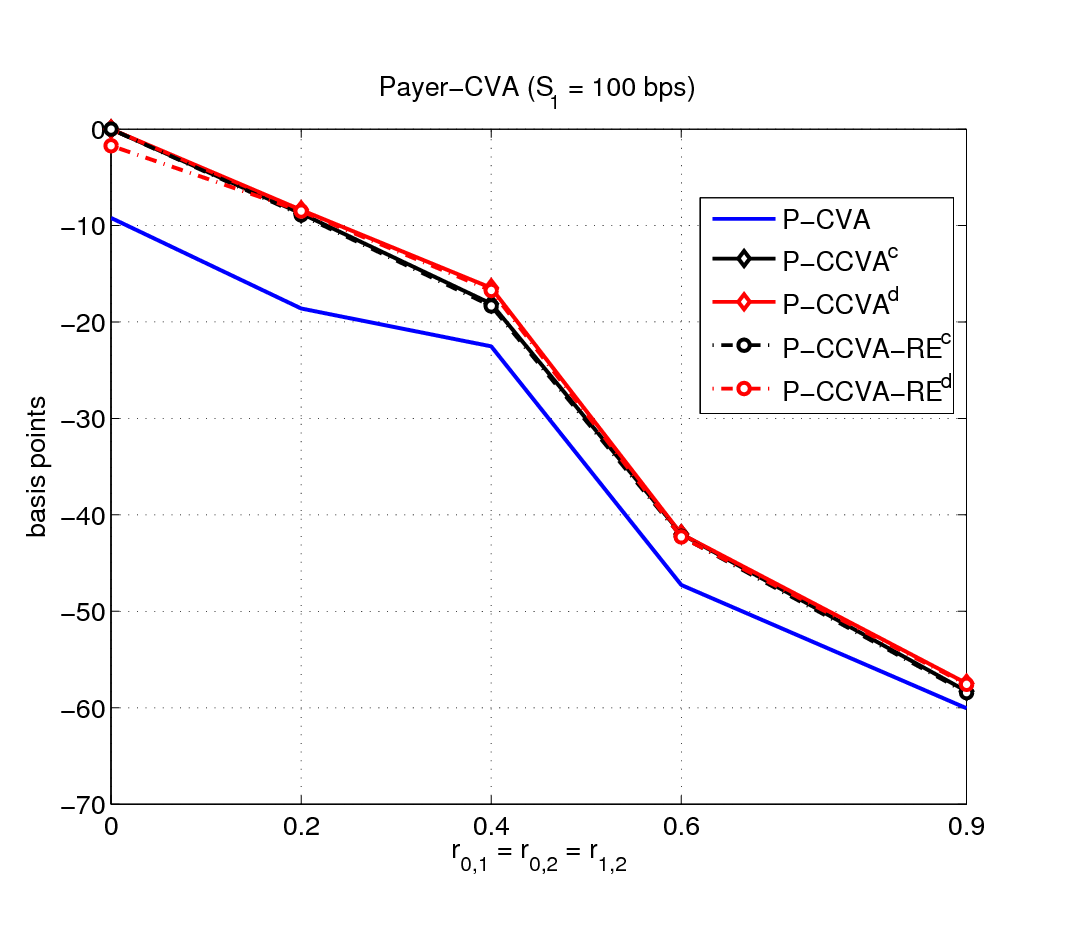} 
\end{center}
\caption{CVA after collateralization (Gap Risk) for a CDS}\label{fig:cvacds}
\end{figure}

Figure \ref{fig:cvacds} refers to a payer CDS contract as underlying. 
If the investor holds a payer CDS, he is buying protection from the counterparty, i.e. he is a protection buyer.
We assume that the spread in the fixed leg of the CDS is 100 while the initial equilibrium spread is about 250.
Given that the payer CDS will be positive in most scenarios, when the investor defaults it is quite unlikely that the net present value be in favor of the counterparty.
We then expect the CVA term to be relevant and DVA negligible, given that the related option will be mostly in the money. This is confirmed by our outputs.
We see in the figure a relevant CVA component (part of the bilateral DVA - CVA)
starting at 10 and ending up at 60 bps when under high correlation.

\item[Q:] This is definitely sizeable

\item[A:] We also see that, for zero correlation, collateralization
succeeds in completely removing CVA, which goes from 10 to 0 basis points.

However, collateralization seems to become less effective as default
dependence grows, in that collateralized and uncollateralized CVA become closer and closer, and for high correlations we still get 60 basis points of CVA, even under collateralization.

\item[Q:] So collateral isn't doing anything here with high default correlation between CDS and counterparty? But why?

\item[A:] The reason for this is the instantaneous default contagion that, under
positive dependency, pushes up the intensity/credit spread of the survived entities, as soon as there is a default of the counterparty.

\item[Q:] Ah...

\item[A:]  You see? The term structure of the on-default survival probabilities (see paper) lies significantly below the one of the pre-default survival probabilities conditioned on pre-default information, especially for large default correlation.

The result is that the default leg of the CDS will increase in value due to contagion, and instantaneously the Payer CDS will be worth more. This will instantly increase the loss to the investor, and most of the CVA value will come from this jump.

\item[Q:] And given the instantaneous nature of the jump, the value at default will be quite different from the value at the last date of collateral posting, before the jump, and this explains the limited effectiveness of collateral under significantly positive default dependence. I see.

\item[A:] This is a pathological case. In general gap risk  will generate a residual Credit Valuation Adjustment (CVA) term after collateralization. Of course you also include the case where you default first, and this will lead to a Debit Valuation Adjustment (DVA) term, also inclusive of collateral netting. You basically have to distinguish eight cases depending on who defaults first, on the sign of the exposure at default, ie whether the deal is in your favour or in favour of the counterparty at the time of default, and finally on the sign of the collateral account. Is the account guaranteeing you at the time of default of your counterparty or is there collateral you posted in the account? The latter may happen with sudden mark to market turns and would be the worst case, as you risk losing both your mark to market (exposure) and your collateral.

\item[Q:] I see, and this generates the familiar CVA and DVA terms after collateralization, that are usually small except under strong gap risk, that we discussed one year ago \cite{Brigo2011faq} and your example just clarified this very nicely. 

\item[A:] Correct, and now you finally need to add the funding costs of maintaining your hedge. This is a sort of carry cost of maintaining the risky assets accounts and also the cash account necessary to keep the hedge going. If you work with your treasury you have your bank treasury interest to pay on the amounts you borrow, and if you spare some borrowing you have a benefit. If you borrow or lend risky assets in a Repo or Stock lending market the associated rates will be included also in the funding cost calculation. 

\item[Q:] Ok so we listed the following fundamental product cash flows and adjustments cash flows:
\begin{itemize}
\item Basic instruments cash flows and coupons before credit collateral and funding
\item Costs and benefits of maintaining the collateralization process (interest on collateral)
\item At default, if before the end of the deal, default closeout cost, with residual CVA and DVA after collateralization;
\item Funding costs and benefits for the cash and risky assets components of the hedge, whether these are obtained by the treasury or directly in the market;
\end{itemize}

\end{itemize}

\section{Funding costs: nonlinearities and aggregation}

\begin{itemize}

\item[A:] Excellent, you have been paying attention. Now, the big difficulty is this. The fact that borrowing and lending happens at different rates, is quite crucial.
So the rate you will have to use to accrue your benefit or cost depends on the future sign of your accounts, because the sign determines whether you are borrowing or lending. Now the future sign of your accounts depends on the future value of your portfolio, which in turns depends on the funding rates depending on that. So you have  a sort of recursive valuation problem. If you write it in terms of discounted cash flows, you find that the risk free rate disappears and that the collateral and funding rates you need to discount and accrue do depend on the future value of the claim you are pricing. Let me repeat this:

\bigskip

{\emph{Consider the collateral and funding rates you need to discount and accrue your product future cash flows in the risk neutral expectation leading to the product price.  These rates depend on the future value of the very same product you are pricing.}}
\bigskip

Hence to determine the value of the claim at time 0, say $V_0$, you need to take an expectation of cash flows accrued and discounted at funding and collateral rates that depend on future $V_t$'s ($t \ge 0$). This is clearly a recursive problem.

\item[Q:] And you mentioned also non-linearity??

\item[A:] Indeed. You see, to better understand this we need to take a step back to classic Black Scholes or risk neutral valuation a la Harrison and Kreps / Pliska. Note that no arbitrage could be based on a different notion of inequalities or betting quotients, see Dalang Morton Willinger or, much earlier, Bruno de Finetti or Frank Ramsey, but we won't deal with that here and we will use the replication approach... 

\item[Q:] My understanding for the importance of this distinction is about zero...

\item[A:]  ... In Black Scholes one uses the self-financing condition of the hedge portfolio plus the modified chain rule (Ito's formula) on the claim value as a function of the risky underlying of the trade to obtain a linear (parabolic) PDE whose solution is the claim price. 

\item[Q:] Ok this is one of the classical derivations of the Black Scholes PDE. 

\item[A:] Now there is the Feynman Kac theorem: solutions of linear (parabolic) PDEs of the Black Scholes type can be written as Expected values in risk neutral pricing of the product final cash flows, discounted at the risk free rate. Viceversa, expected values in the risk neutral measure can be written as solutions of linear PDEs. This is, incidentally, how one can solve PDEs by Monte Carlo simulation.

\item[Q:] Is that why quantitative finance has largely abandoned the PDE approach in favour of stochastic methods? Because in high dimension simulation is possible while numerical PDE solution in dimension larger than 3 is very hard?

\item[A:] Partly, but there is also the appealing risk neutral interpretation and the measure Q that we discussed one year ago \cite{Brigo2011faq}, a measure under which all risky assets have the risk free return $r$, remember? This is an excellent model for a fair market...

\item[Q:] We digress... Ok, in Black Scholes the self financing condition plus Ito's formula gives me a linear PDE that I could also solve by simulation as an expected value of discounted cash flows under a risk neutral measure, discounted at the risk free rate. Got that. So what? 

\end{itemize}

\section{Nonlinear PDEs, nonlinear expectations and BSDEs}

\begin{itemize}

\item[A:] What? Well, now imagine a slightly different situation, that is what happens in our funding case with asymmetric borrowing and lending rates. If the discount and accruing rates depend on the future values of the very same solution we are trying to compute through the discounted expectation, this does not translate into a linear parabolic PDE as in the Black Scholes theory, but rather into a non-linear PDE.

\item[Q:] That's where the non-linearity comes from, I see... in the Black Scholes theory the risk free discount and accruing rate does not depend on the future value of the claim, so I have a straightforward expectation... with funding costs I don't... What are the consequences of this nonlinearity? It does not look good from your face...

\item[A:] Is my makeup melting? No matter. It is indeed worrying. As you probably  guessed, the valuation of a portfolio is aggregation dependent. In the classical pricing theory a la Black Scholes, if we have 2 or more derivatives in a portfolio we can price each separately and then add up, to obtain the whole price of the portfolio. Moreover,  while in Black Scholes the price of the whole portfolio to one party is minus the price of the portfolio to the other party, with funding costs this symmetry is broken, because the calculation is not bilateral in general.

\item[Q:] Why not?

\item[A:] Because we don't know the other entity funding policy and hence we cannot know their funding costs. And why should they accept to pay our funding costs for us and why should we accept to pay them theirs? After all, their funding inefficiency is not our fault... however we have to be careful because there are two symmetries that are violated here and it's easy to confuse them.

\item[Q:] Well now you were talking about symmetry with respect to the two trading parties I guess. If B trades with C and sees a price $V$, then $C$ should see a price $-V$ in the same trade with B. This is the first symmetry, and it is violated by funding costs because one party cannot check the funding costs of the other one easily, so that charging each other funding costs explicitly and agreeing on the other party charge is not possible. What is the other symmetry?

\item[A:] It's the long-short symmetry. Without credit, collateral and funding risk, if you are long a product with discounted cash flows $\Pi$, the price is the opposite of the short position with cash flows $-\Pi$. This symmetry is broken also via the nonlinearity of funding costs.

\item[Q:] But probably that's embedded in the quoted price without us realizing it and it becomes part of the bid offer negotiation?

\item[A:] Possibly, but some banks do charge the funding inclusive price to the client anyway by quoting an explicit funding component, or "FVA". However, let's go back to nonlinearities. 

\item[Q:] OK

\item[A:] Because of nonlinearity, these aspects of additive decomposition no longer hold in presence of funding costs. The price becomes aggregation dependent. You need to decide a priori at what portfolio level you are pricing because prices of sub-portfolios do not add up. And once the level of aggregation is set, the funding valuation problem is non-separable, because the funding rates patch everything together by depending on the future value of the whole portfolio at the chosen aggregation level, and not just of its local components. A holistic approach is needed and consistent modeling across trading desks and asset classes is needed.   Internal competition in banks does not favour this but it is a must if one is to try and price/risk manage these aspects consistently.  

\item[Q:] Sounds revolutionary in scope, even if it comes from a very simple financial fact, borrowing and lending at different rates...

\item[A:] Well the maths is not revolutionary, this kind of problem with asymmetric rates had been explored way back already (without credit risk or collateral) by El Karoui, Peng and Quenez \cite{elkaroui}, and the resulting tool, Backwards Stochastic Differential Equations (BSDEs), more on these in a minute, have been know for a long time in finance. In fact if we had to give credit where credit is due, to the best of my knowledge the first to come up with the understanding that different discount rates for borrowing and lending would lead to fundamental nonlinearities was Bergman back in 1995 \cite{bergman85}. The new aspect is that now these nonlinear features are showing up very fundamentally, much more so than in the past, and so it is very hard to ignore them even for mainstream quantitative analysts. 

\item[Q:] That's a lot to take in!

\end{itemize}

\section{Funding costs: Claim dependent pricing measures}

\begin{itemize}

\item[A:] I'm not done yet. The classical transaction-independent arbitrage free price is lost, now the price depends on the specific entities trading the product and on their policies.
In fact the so called pricing measure now depends on the specific trading strategy adopted for the deal being priced. This is a point that is stressed well both in Pallavicini et al \cite{BrigoPallaPerini} \cite{BrigoPallaPerini2}, and in \cite{RutkowskiFunding}. In Pallavicini et al the risk neutral expectations such as $\mathbb{E}^f$ are taken under a specific pricing measure depending on the funding rates $f$ of the treasury, that in turn depend on the future values of the claim we are pricing. Hence the pricing measure depends on the claim we are pricing, as I explained earlier.  I'm stressing this point many times because it is key. It is like "relativizing" the pricing measure, that is no longer the measure associated with the abstract and somewhat fictional but universal risk free bank account $B$, see again our previous conversation \cite{Brigo2011faq}.

\item[Q:] So a very simple financial fact like the fact that borrowing and lending may happen at different rates is generating all this amount of complications and nonlinearity?

\item[A:] Correct. In fact I mentioned expectations above but a better term would be "nonlinear expectations", namely expectations under a probability measure that sort of depends on the random variable whose expectation we are taking. Another way of saying this is that nonlinear PDEs do not translate properly in expected values. That is the basic Feynman Kac theorem but it only works for risk neutral valuation without funding costs. In fact, nonlinear PDEs translate probabilistically into BSDEs. BSDEs may be difficult animals to solve numerically, but they have been around a long time, especially in their semi- and quasi-linear variants. They have been used also for no good deal theory. There are very technical works on nonlinear expectations, see for example \cite{pengnonlin}.

\item[Q:] Ok, it's a pity I hadn't realized the importance of this and had thought these to be technicalities.

\item[A:] It's quite important. You could see it partly as the end of Platonic pricing. There is no Platonic measure Q in the sky to price all derivatives with an expectation where all assets have the risk free return $r$. 


Now the pricing measure is product dependent, and every trade will have a specific measure. This is an implication of the PDE non-linearity and we need to decide the level of aggregation a priori. 

\item[Q:] Mmmhhh

\item[A:] Other markets had realized a long time ago that a product price would also depend on the conditions under which the product itself is traded and on the company policies. 
Finance arrived at this conclusion quite late, even if market practitioners had been doing this in a sort of implicit way. 

\item[Q:] Well I think we need to agree on the term "price". I tend to see "price" as something coming out from a process of bid and offer, and negotiation, whereas there is another term I'd use, "value", for the actual value I perceive in entering a trade. The latter is not necessarily what I pay for it.

\item[A:] Well of course the distinction between price and value is a classic in economics, but in the end does it matter?

\item[Q:] I think it does. You mentioned that charging "FVA" to the counterparty is controversial, and that FVA cannot be bilateral, since we do not know the funding policy of our counterparties. So even if DVA was giving us some hope to realign symmetry of prices after introducing CVA, funding finally destroys the law of one price and makes prices a matter of perspective if included: direction, aggregation level... So should this be implicitly quoted in the Bid Ask price and be smoothed away by negotiation? Should we think more explicitly to an Equilibrium approach? These are important questions to me.

\item[A:] Ok, I see you are critical (and not without reason) about explicitly charging a (even unilateral) FVA to a counterparty, but if we don't, how would you use the funding inclusive price? Should we avoid computing it in the first place?

\item[Q:] Again, is the funding inclusive "price" a real price? 
Each entity computes a different funding adjusted price for the same product and at different level of aggregation. The funding adjusted "price" is not a price in the conventional sense. We may use it  for cost/profitability analysis or to pay our treasury, but can we charge it to  a client? Again, why should the client pay for our funding inefficiencies? Indeed, it is more a "value" than a "price".

\item[A:] However you call it, it is hard to compute quickly.

\item[Q:] Even if we decided to compute it for internal cost analysis or profitability analysis, we woud have to deal with BSDEs, nonlinear PDEs, how can one make these things work in practice?

\item[A:] It's quite hard. 
Banks may force separability between FVA and CVA, DVA etc and additivity of the different adjustments by assuming the borrowing and lending rates to be symmetric, perhaps an average of the two, and by linearizing payoffs and simplifying discounting, assuming statistical independence, obtaining that funding costs reduce simply to different discount adjustments. Of course this only holds under heroic assumptions and is not precise at all, and in specific situation can be quite misleading.

\item[Q:]  I see... that's why most people in the industry address funding through discounting. Funny I had never read the complete story.
 
\item[A:] This is hardly complete, not least because it is not obvious at all that funding costs are really replicable, so that the derivations above, based on replication / hedging arguments, may fail and more actuarial based approaches could be taken. This is currently under investigation in \cite{brigodurand} from what I heard at a seminar here in London.

\item[Q:] I have to say also that now I understand better the point on non-separability you made one year ago \cite{Brigo2011faq} [Sighing].  A number of practitioners advocate a ``Funding Valuation Adjustment", or FVA, that would be additive so that the total price of the portfolio would be

RISK FREE PRICE + DVA - CVA - FVA

However, we have seen that proper inclusion of funding costs leads to a recursive pricing problem where the discount and accrual rates in the so called FVA term depend on the whole future value of the claim, including future CVA and DVA, so that the pure decomposition of the risks above is delusional, FVA depends on all other risks as well and is not a pure funding adjustment.

\item[A:] Good to hear that.

\item[Q:] I'm frustrated from the fact that the literature on funding is quite sparse and partial in many cases.

\item[A:] Well it is developing fast, but we may review a little the past contributions if that helps you put all we said into context?

\item[Q:] Yes that would be good, thank you.

\end{itemize}

\section{Funding costs: earlier contributions}

\begin{itemize}

\item[A:] Ok let's take a couple of steps back and look at some of the earlier works. They are partial but still quite important.

\item[Q:] For example?

\item[A:] You may recall that the industry paper \cite{Piter2010} considers the problem of replication of derivative transactions under collateralization but without default risk and in a purely classical Black and Scholes framework, considering then two relatively basic special cases. When you derive the first basic results that way, you have to be very careful on  the way you formulate the self financing condition, see for example \cite{brigoliu2012}. 

\item[Q:] Ok but aside from the self financing condition, I would imagine that default modeling is important in collateral and funding, isn't it? That is the reason why collateral is introduced in the first place.

\item[A:] Indeed, but sometimes explicit modeling of default is avoided because inclusion of default risk is quite complicated. The fundamental funding implications in presence of default risk have been considered in simple settings first in \cite{MoriniPrampolini}, see also \cite{Castagna2011}. These works focus on particularly simple products, such as zero coupon bonds or loans, in order to highlight some essential features of funding costs.  \cite{Fujii2010} analyzes implications of currency risk for collateral modeling, and has been followed by a similar paper later in \cite{Piterbarg2012}. \cite{Burgard2011} resorts to a PDE approach to funding costs. 

\item[Q:] That's quite a number of references already. What is the level of generality here and what are the most comprehensive approaches?

\item[A:] As I mentioned above Crepey (2012) \cite{Crepey2012a} with \cite{Crepey2012b}, and Pallavicini et al (2011, 2013) \cite{BrigoPallaPerini} with \cite{BrigoPallaPerini2} remain the most general treatments of funding costs to date. These papers show how complicated it is to formulate a proper general framework with collateral and funding costs but inclusive of default risk.  In fact, more recently, \cite{RutkowskiFunding} did a good job of expressing the funding costs valuation problem in mainstream asset pricing and replication theory, but with a high technical level of complexity and a number of assumptions. The end result is very similar to \cite{BrigoPallaPerini} and \cite{BrigoPallaPerini2} in terms of pricing implications, aggregation, strategy dependent measures, nonlinearities... 

\item[Q:] Good, we discussed above some of these  general frameworks, but  what are instead the findings in Morini and Prampolini \cite{MoriniPrampolini}? I heard of this paper when it was still a preprint. 

\item[A:] One important point in Morini and Prampolini \cite{MoriniPrampolini} is that in simple payoffs such as bonds DVA can be interpreted as funding, in order to avoid double counting. However, this result does not extend to general payoffs, where different aspects interact in a more complex way and the general approaches we described above are needed.

\end{itemize}

\section{Funding costs: DVA and FVA}

\begin{itemize}

\item[Q:] What is the basic argument to interpret DVA in Funding terms?

\item[A:] DVA is related to funding costs in particular situations when the payout is uni-directional, e.g. shorting/issuing a bond, borrowing in a loan, or going short a call or put option. 
Indeed, if we are short simple products that are uni-directional in terms of payments to happen in the future, then we are basically borrowing. 

\item[Q:] I see that. 

\item[A:] As we shorted a bond or a call option, for example, we received cash equivalent to the product price in the beginning, say $V_0$, and we will have to pay the product payout in the end. 
This cash $V_0$ we received as product price can be used by us to fund other activities, and allows us to spare the costs of funding this same amount of  cash $V_0$ from our treasury.

\item[Q:] All right, now if I try to compute the interest amount I am saving by not borrowing an amount $V_0$ from the treasury...

\item[A:] Indeed, how much would the treasury charge your trading desk if you borrowed that cash $V_0$? Our treasury usually funds in the market, and the market charges our treasury a cost of funding that is related to the borrowed amount $V_0$, to the period $T$ on which we are borrowing, and to our own bank credit risk. 

\item[Q:] I see, these are roughly the same drivers as the drivers of DVA, that depends on our own bank default risk and on the price we would be trading without credit risk. 
In this sense the funding cost we are sparing when we avoid borrowing looks similar to DVA: it is related to the price of the object we are shorting and to our own credit risk. I see that now. But...

\item[A:] You are going to argue that there are a  number of assumptions that are needed to identify DVA with a funding benefit, and indeed this is the case. The single direction flows is one of them, but there is a number of other points. Your treasury department for example does not necessarily fund single trades separately but often adopts a strategy based on internal fund transfer pricing, maturity transformation, netting sets, to a point where the rate that is charged to the trading desk is hardly a simple credit spread of the bank over the risk free rate. In this sense the actual funding cost is a function of the treasury policy and may be more complicated to analyze. Pallavicini et al \cite{BrigoPallaPerini} \cite{BrigoPallaPerini2} look at this and give two examples of treasury funding policy, showing how valuation changes according to those and the simple DVA analogy does not work.

\item[Q:] All right, ten more papers to read... 

\end{itemize}

\section{CCPs}

\begin{itemize}

\item[Q:] You know, all these matters with funding make me wish there was a way to avoid this completely. Last year in our earlier conversation \cite{Brigo2011faq} we talked about margin lending. Thinking back to what we had said,  do you think there are many reasons why one might prefer to trade through a margin lender than through central clearing, for example a Central Clearing Counterparty (CCP) like DTCC, LCH, CME...

\item[A:] Ah, CCPs, they are like the elephant in the room [smiles tiredly]... Well you see, sometimes there are particular trades that are sensitive and that the bank management would not have appearing in central clearing-houses, or other times the trades are too specific and non-standard, so they would be a little outside the usual CCP domain. 

\item[Q:] But for those that are relatively standard? After all, we don't see many exotic products these days.

\item[A:] It's not just the type of product, it may have to do with the specific environment or context for a trade. But now it's quite late [watches clock on the wall], let's call it a day and continue tomorrow, even quants need to eat and rest!

\item[Q:] [Yawns] In fact I'm exhausted, this discussion is exciting but also very tiring, I'll dream of collateral damage, wrong way lanes, probability measures and CCCP!!!

\item[A:] [Stretching] I'll dream about a world withoug scarce resources and without economics, money and finance, perhaps something like Star Trek.

\item[Q:] [One day later...] All right... let's talk about CCPs a little more in detail. Why do we care about CVA etc if CCPs will kick in?

\end{itemize}

\section{CCPs: are CVA and Funding analysis still relevant?}

\begin{itemize}

\item[A:] A common misunderstanding in the industry and academia is that the onset of CCPs will make all calculations of counterparty risk pricing and funding costs irrelevant.

\item[Q:] And that is not true? 

\item[A:] I can only tell you what I believe. I think this is indeed a misunderstanding and we need to move beyond this narrow understanding of valuation, hedging, credit and liquidity risk and collateral modeling. 

\item[Q:] What do you mean by "narrow"?

\item[A:] Let us start with an obvious question: what are CCPs? CCPs are commercial entities, and I point out "commercial", that, ideally, would interpose themselves between the two parties in a trade.

\item[Q:] Ok, I think I know this
 
\item[A:]
Consider the following points [enumerating with fingers]:
\begin{itemize}
\item Each party will post collateral margins say daily, every time the mark-to-market goes against that party. This is called "variation margin". 
\item Collateral will be passed by the CCP to the other party as a guarantee. 
\item If a party in the deal defaults and the mark-to-market is in favour of the other party, then the surviving party has obtained the collateral from the CCP and will not be affected, in principle, by counterparty risk. 
\item Moreover, there is also an initial margin that is supposed to cover for additional risks like deteriorating quality of collateral, gap risk, wrong way risk, and so on.
\item This initial margin is kept by the CCP and not passed to the opposite party. The CCP may charge both parties for the initial margin. 
\item If the CCP defaults initial margins can be lost 
\end{itemize}

\item[Q:] Ah but CCPs cannot default, come on, that's the whole point.

\end{itemize}

\section{CCPs: Past Defaults}

\begin{itemize}

\item[A:] On the contrary, CCPs can default and did default. Defaulted ones:
\begin{itemize}
\item 1974: Caisse de Liquidation des Affaires en Marchandises; 
\item 1983: Kuala Lumpur Commodity Clearing House; 
\item 1987: Hong Kong Futures Exchange. 
\item The ones that were close to default- 1987: CME  and OCC, USA; 1999: BM{\&}F, Brazil.
\end{itemize}

\item[Q:] Seriously, how can you expect entities like LCH or DTCC to default? They are too big to.. oh my...

\item[A:] Too big to fail? Indeed, if the CCPs become aggregated more and more into a few very large ones, they could become the mother of all "too big to fail". See \cite{Miller2011} for example, dealing also with conflict of interests in CCPs. 

\item[Q:]  Ok that is certainly important and alarming, but we were talking about valuation?

\end{itemize}

\section{CCPs: Netting benefits}

\begin{itemize}

\item[A:] I think we need to get the big picture straight before going into details. So back to that: the benefit of 
CCPs is supposed to be the large netting they may introduce in the financial system. Let me illustrate that with a drawing [gets up, draws Figure \ref{fig:kiff} on the whiteboard].

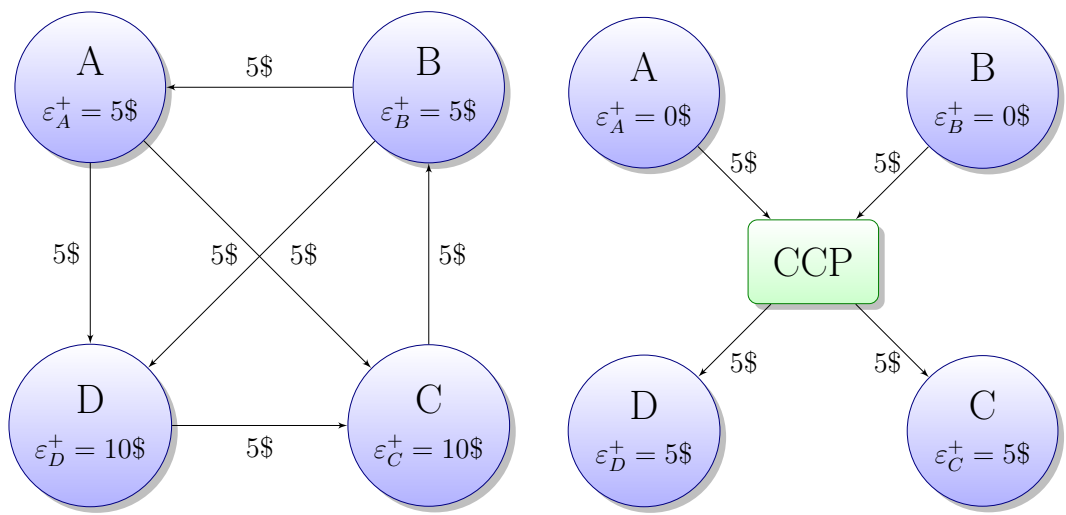
\begin{figure}
\begin{center}
\scalebox{0.9}{
\begin{tikzpicture}
   \node [trader, node distance=5cm] (A) {\shortstack[cc]{{\Large A}\medskip\\$\varepsilon_A^+=5\$$}};
   \node [trader, right of=A, node distance=5cm] (B) {\shortstack[cc]{{\Large B}\medskip\\$\varepsilon_B^+=5\$$}};
   \node [trader, below of=B, node distance=5cm] (C) {\shortstack[cc]{{\Large C}\medskip\\$\varepsilon_C^+=10\$$}};
   \node [trader, below of=A, node distance=5cm] (D) {\shortstack[cc]{{\Large D}\medskip\\$\varepsilon_D^+=10\$$}};
   \path [from] (A) -- node [above] {$5\$$} (B);
   \path [to] (A) -- node [left] {$5\$$~~} (C);
   \path [to] (A) -- node [left] {$5\$$} (D);
   \path [from] (B) -- node [right] {$5\$$} (C);
   \path [to] (B) -- node [right] {~~$5\$$} (D);
   \path [from] (C) -- node [below] {$5\$$} (D);
\end{tikzpicture}}~~~~~~
\scalebox{0.9}{
\begin{tikzpicture}
   \node [trader, node distance=5cm] (A) {\shortstack[cc]{{\Large A}\medskip\\$\varepsilon^+_A=0\$$}};
   \node [trader, right of=A, node distance=5cm] (B) {\shortstack[cc]{{\Large B}\medskip\\$\varepsilon^+_B=0\$$}};
   \node [trader, below of=B, node distance=5cm] (C) {\shortstack[cc]{{\Large C}\medskip\\$\varepsilon^+_C=5\$$}};
   \node [trader, below of=A, node distance=5cm] (D) {\shortstack[cc]{{\Large D}\medskip\\$\varepsilon^+_D=5\$$}};
   \node [market, below right of=A, node distance=3.535cm] (X) {{\Large CCP}};
   \path [to] (A) -- node [above] {~~$5\$$} (X);
   \path [to] (B) -- node [above] {$5\$$~~} (X);
   \path [from] (C) -- node [below] {$5\$$~~} (X);
   \path [from] (D) -- node [below] {~~$5\$$} (X);
\end{tikzpicture}}
\end{center}
\caption{Bilateral trades and exposures without CCPs (on the left) and with CPPs (on the right). Each node lists the sum of positive exposures, each arrows the due cash flows. The diagram refers to the discussion in \cite{Kiff2009}.\label{fig:kiff}}
\end{figure}

\item[Q:] Ok, even as a toy example this definitely looks like a good thing. However I hardly expect this to cover all the macro implications?

\item[A:] There are further important considerations on the macroeconomic and systemic effects of CCPs. As discussed in \cite{Piron2012}, it is worth keeping in mind that CCPs are usually highly capitalized. All clearing members post collateral in an asymmetric way. Initial margin means clearing members are over-collateralized all the time. 

\item[Q:] Do you think this initial margin requirements will bring about a liquidity strain?

\item[A:] Well in a way you may see Collateral and Margining as a machinery that transforms partly credit risk into liquidity risk. 

\item[Q:] Say that again?

\item[A:] As we discussed above, collateral greatly reduces counterparty risk, even if one has to be careful with contagion, gap risk, disputes, collateral volatility, FX effects etc. Ok? So overall in non-pathological situations collateral reduces counterparty credit risk almost to zero, but forces you to find funds to post the collateral itself if you are the party who is facing adverse mark to market. 

\item[Q:] Ah I see...


\item[A:] Where will you find liquid assets or cash to be posted in the collateral account? This may prove to be a liquidity strain as you need to guarantee collateral is there at every margin call.  On top of variation margin calls, you also have initial margins or independent amounts, depending on the context, covering the ``pathological" risks such as contagion, gap risk, extreme wrong way risk etc. The overall effect is that in most situations one will not face relevant credit risk any more but there is the need to find liquidity to keep the initial and variation margins  payments going. 

\item[Q:] So you are saying this initial margin requirements will bring about a liquidity strain, either through CCPs or SCSAs?

\end{itemize}

\section{CCPs: Liquidity Strain}

\begin{itemize}

\item[A:] The TABB Group \cite{Tabb2011} says extra collateral impact of full CCP onset could be about 2 $\$$ Trillion!! It's about the GDP of the UK so we are definitely talking about a big liquidity impact on the system

\item[Q:] But at least the netting benefit we discussed above, which is key, is hardly debatable, right? 

\item[A:] Looking at the systemic implications, netting should improve in principle as everything goes through CCPs, as we have seen in the toy example above. However, in reality this is not so clear. A typical bank may have a quite large number of outstanding trades, making the netting clause quite material. With just one CCP for all asset classes across countries and continents, netting efficiency would certainly improve. 

\item[Q:]  And it's not like that in reality?

\item[A:] In real life CCPs deal with specific asset classes or geographical areas, and this may even reduce netting efficiency compared to now, as discussed in \cite{Rama2013}. CCPs compete with each other and one can be competitive in specific areas but hardly in all of them. Some CCPs will be profitable in specific asset classes and countries. They will deal mostly with standardized transactions. Even if CCPs could function across countries, bankruptcy laws can make collateral held in one place unusable to cover losses in other places, as pointed out in \cite{Singh2011b}.

\item[Q:] Are you saying that there is a lack of consistency and coordination across jurisdictions?

\end{itemize}

\section{CCPs: Jurisdictions}

\begin{itemize}

\item[A:] The geographical angle seems to be an issue, with no international law addressing how CCPs would connect through EMIR/CDR 4/Basel III and DFA. For example, as of 2012 there was ``No legal construct to satisfy both Dodd Frank Act and EMIR and allow EU clients to access non-EU CCPs" \cite{Wayne2012}. 

\item[Q:] Could this really make a difference? What else can make CCPs risky?

\item[A:] To compete with each other CCPs may lower margin requirements, which would make them riskier, remembering also the above CCPs defaults. In the US, where the OTC derivatives market is going through slightly more than 10 large dealers and is largely concentrated among 5, we could have a conflict of interest. If CCPs end up incorporating most trades currently occurring OTC bilaterally, then CCPs could become ``too big to fail'' \cite{Miller2011}, as we mentioned earlier. 

\item[Q:] But despite all this macro discussion you cannot deny that CCPs would greatly improve credit risk of the system?

\item[A:] Well I think the ``too big to fail" danger cannot be ignored. 

\item[Q:] Fair enough, I'd agree with that...

\end{itemize}

\section{CCPs: Initial margins potential loss and CCPs competition}

\begin{itemize}

\item[A:] ... and consider also the following. The CCP does not post part of the collateral (initial margins) directly to the entities trading with it, as the collateral agreement is not symmetric. We mentioned this above already. If you trade with collateral but without a CCP, say through a ISDA SCSA and subsequent updates, you are in a situation like this [draws Figure \ref{fig:bilateral} on the whiteboard]. So you exchange both variation and initial margins. 

\item[Q:] Ok, and what is different with CCPs?

\item[A:] With CCPs, one has the over-collateralization cost to lose in a uncollateralized type CVA situation. The situation is like this [draws Figure \ref{fig:clearing} on the board]. The CCPs passes the variation margin to the other party but not the initial margin. 
Hopefully, the default probability of the CCP itself is low, making CVA towards the CCP small, bar strong contagion, gap risk and wrong-way risk, that may or may not be adequately covered by initial margins.

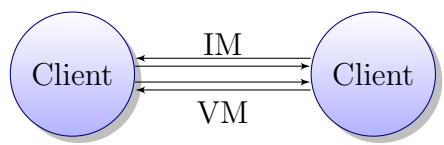
\begin{figure}
\begin{center}
\scalebox{1}{
\begin{tikzpicture}
   \node [trader, node distance=4cm] (D1) {Client};
   \node [trader, right of=D1, node distance=4cm] (D2) {Client};
   \path [to,transform canvas={yshift=3pt}] (D1) -- node [above] {IM} (D2);
   \path [from,transform canvas={yshift=6pt}] (D1) -- node [above] {} (D2);
   \path [to,transform canvas={yshift=-3pt}] (D1) -- node [below] {} (D2);
   \path [from,transform canvas={yshift=-6pt}] (D1) -- node [below] {VM} (D2);
\end{tikzpicture}}
\end{center}
\caption{Flows of initial and variation margin within counterparties acting in a bilateral contract via a CSA. IM stands for initial margin, while VM for variation margin. The two margins flow in both directions. }
\label{fig:bilateral}
\end{figure}

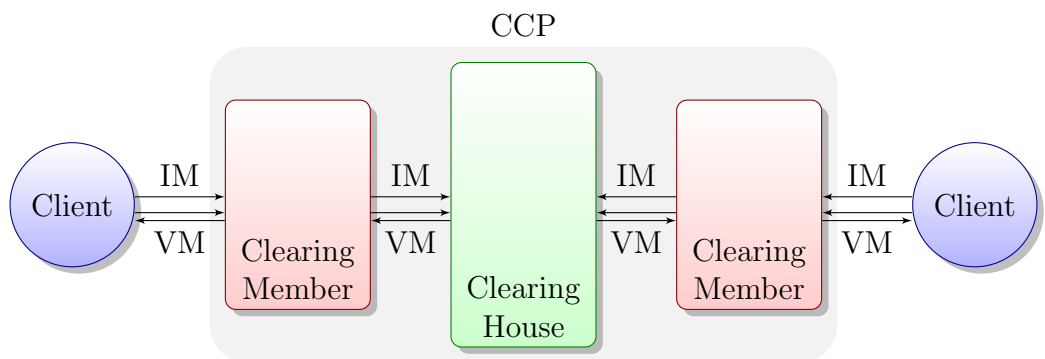
\begin{figure}
\begin{center}
\scalebox{1}{
\begin{tikzpicture}
   \node [trader, node distance=3cm] (D1) {Client};
   \node [bank, right of=D1, text height=2cm,node distance = 3cm] (CM1) {Clearing Member};
   \node [market, right of=CM1, text height=3cm, node distance = 3cm] (CCP) {Clearing House};
   \node [bank, right of=CCP, text height=2cm,node distance = 3cm] (CM2) {Clearing Member};
   \node [trader, right of=CM2, node distance=3cm] (D2) {Client};
   \path [to,transform canvas={yshift=3pt}] (D1) -- node [above] {IM} (CM1);
   \path [to,transform canvas={yshift=-3pt}] (D1) -- node [below] {} (CM1);
   \path [from,transform canvas={yshift=-6pt}] (D1) -- node [below] {VM} (CM1);
   \path [to,transform canvas={yshift=3pt}] (CM1) -- node [above] {IM} (CCP);
   \path [to,transform canvas={yshift=-3pt}] (CM1) -- node [below] {} (CCP);
   \path [from,transform canvas={yshift=-6pt}] (CM1) -- node [below] {VM} (CCP);
   \path [from,transform canvas={yshift=3pt}] (CM2) -- node [above] {IM} (D2);
   \path [to,transform canvas={yshift=-3pt}] (D2) -- node [below] {} (CM2);
   \path [from,transform canvas={yshift=-6pt}] (D2) -- node [below] {VM} (CM2);
   \path [from,transform canvas={yshift=3pt}] (CCP) -- node [above] {IM} (CM2);
   \path [to,transform canvas={yshift=-3pt}] (CM2) -- node [below] {} (CCP);
   \path [from,transform canvas={yshift=-6pt}] (CM2) -- node [below] {VM} (CCP);
   \begin{pgfonlayer}{background}
   \node [background, fit=(CM1) (CM2) (CCP), label=above:CCP] {};
   \end{pgfonlayer}
\end{tikzpicture}}
\end{center}
\caption{Flows of initial and variation margin within counterparties acting in a centrally cleared contract. IM stands for initial margin, and it flows only towards the CCP. VM stands for variation margin, and it flows in both directions. }
\label{fig:clearing}
\end{figure}

\item[Q:] Ok but what are the implications on valuation and analytics? As I mentioned earlier, are CCPs safe enough to make all calculations we discussed in the previous meetings pointless?

\item[A:] [Sighs tiredly] Even with CCPs in place, one needs a strong analytical and numerical apparatus for pricing/hedging and risk. For all the reasons we illustrated, CCPs are not the end of CVA and its funding costs extensions. We need to consider and analyze:
\begin{itemize}
\item initial margin charges across different CCPs to see which ones best reflect actual gap risk and contagion. This requires a strong and consistent pricing apparatus;
\item computing counterparty risk associated with the default of the CCP itself and of clearing members, taking into account backup members;

\item understanding quantitatively the consequences of the lack of coordination among CCPs across different countries and currencies. 
\end{itemize}
These are just a few examples of the type of valuation problems on has with CCPs. 

\item[Q:] Look, I know you are doing your best and you look tired now, but I have to be honest here, I am getting skeptical...

\item[A:] On what? Nothing wrong with that however, in general, as long as you motivate your skepticism convincingly

\item[Q:] Here, please have some tea, and let me explain. Let's take it easy. 
You seem to be saying that we may want to compare different margin proposals from different CCPs to see which ones best reflect risk pricing, and to make sure the CCP is careful 
enough? Forgive me but this does not look realistic to me. 

\item[A:] [Sipping tea] Mmmhhh I think I know what you are going to say next...

\item[Q:] Allow me to say it anyway. Most banks would trade with the CCP offering good conditions rather than with a CCP with higher costs. What types of market players are going to accept a higher cost because it agrees with their own risk analysis and they think that paying higher costs makes the CCP, or the system more generally, safer? 

\item[A:] I'll use a loose analogy here, but suppose there is a home insurance provider that is on record for having been close to bankruptcy several times and also known for not offering good coverage in that the insurer tries to dodge most claims. There is a second one that is know to be more solid and that generally has a better record of paying claims. However the second one is more expensive. Which one would you pick to insure your own home?

\item[Q:] In this case I agree I would pick the expensive one, but I don't think this is a good analogy. Are market players resorting to CCPs because they really believe this will cover their risks or are they going to do that just to avoid heavy capital requirements associated with CVA Value at Risk etc? In the former case you may be right, but in the latter this looks more like car insurance where you buy just the basic insurance requested by law, where you make sure the insurance pays the other party damage in case you are guilty in a car accident. In this case you are going for the insurance company with lowest cost, since it is the other party in the accident that will possibly face the low quality of the insurance you bought at a low cost for you.

\item[A:] I see your objection, it is a little cynical but I cannot deny we need to look at this possibility.  However why don't we assume banks are trying to do the right thing for a start and take that as a reasonable basis for our analysis?

\item[Q:] Ok we can reason on this setup, but please keep in mind my caveats, I don't think they are that unrealistic really. But I have another objection...

\item[A:] [rolling eyes skywards...] Let's have it!

\item[Q:] Do you really think a quantitative analysis like the one you mentioned above would not be overridden by political and administrative considerations? The decision to trade with a particular CCP might be driven by existing relationship, by past cooperation, even by peer pressure in some cases. Modeling might have nothing to do with it. 

\item[A:] Of course. However even if you do not ``optimize" over the CCP choice, and you accept one for ``political" reasons, you may want to check the costs the CCP is imposing you to see if they are roughly right in terms of order of magnitude, and you may want to price residual risks. Do you agree with that?

\item[Q:] Ok I can accept that. But if we remove the optimal choice of the CCP,  do the remaining problems require any modeling novelty?

\item[A:] In order to consistently price a deal cleared through a CCP (or alternatively a deal under a bilateral CSA) by taking into account residual credit risk and gap risk, initial and variation margins, collateral, close-out netting rules, and wrong-way risk, one needs to have an analytical set of tools similar to the collateralized CVA/DVA/FVA machinery we described earlier.  So yes, this requires the type of valuation we have been talking about so far. This is investigated in the same framework as CVA, DVA and Funding Costs in \cite{CCPPallaBr2013}.

\item[Q:] Could I have some numerical examples, just to get a feeling, without reading the whole paper with derivations and equations?

\end{itemize}

\section{Valuation with C-D-FVA under CCPs: Numerical Examples with Initial and Variation Margins}

\begin{itemize}

\item[A:] I can show you some figures which I obtained re-implementing the equations in \cite{CCPPallaBr2013} with slightly different inputs. We are talking about a client, say "I", trading with a counterparty via a CCP, and the counterparty "C" here is taken to be the clearing member in the CCP. The investor "I" deals with the clearing member "C" without seeing the real counterparty.

\item[A:] Understood. Shall we look at some figures now?

\item[A:] Good. Let me say that here we are looking at a case where the Investor credit risk is higher than the credit risk of the clearing member, as is realistic to expect. If a clearing member (CM) is riskier than I am as an investor, then I have not much faith in the CCP employing that CM. So I expect myself to be riskier. 

\item[Q:] This is clear.

\item[A:] Fine, let me show you first this [brings on the tablet screen Figure \ref{fig:DVA_qs}].

In all the figures I will bring up I am showing you the pricing for a 10 years receiver Interest Rate Swap (IRS) cleared by a CCP, initially traded at par. We start with funding costs and credit risk disregarded. I show you the DVA impact that is the main credit feature, since I am assuming I am pricing from the point of view of "I", the investor, in a scenario where "I" has much higher credit risk that the CM, so that DVA will dominate CVA. You remember our debate on CVA and DVA last year, right? Even with collateral we'll have DVA as I take a margin period of risk of 5 days. Not so unrealistic since many collateral disputes are settled in about 1 week or ten days. \cite{Brigo2011faq} 

\begin{figure}
\begin{center}
\includegraphics[scale=1]{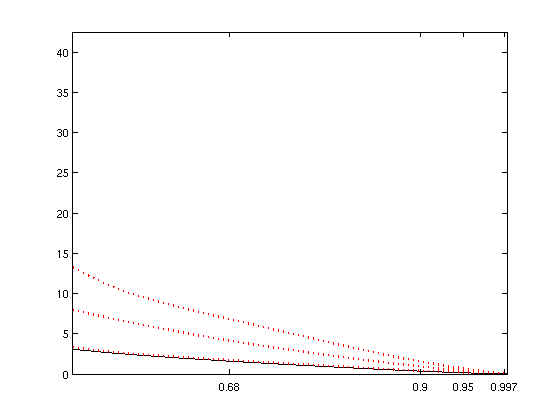} 
\end{center}
\caption{DVA price and percentiles for a 10y receiver IRS cleared by a CCP with initial margins at different levels of safety}\label{fig:DVA_qs}
\end{figure}

\item[Q:] Of course, but what is on the x axis here? 

\item[A:] On the x axis you see the confidence level that is used in the calculation of the initial margin. Basically the initial margin is taken as the Value at Risk of the trade with a confidence level shown in the x axis. Clearly, the higher the confidence level the safer the protection offered by the initial margin, and indeed all graphs are decreasing.

\item[Q:] Ok, what is on the y axis? 

\item[A:] This is  the DVA (lowest continuous line) priced  by the client and a few quantiles of DVA future discounted cash flows whose expectation gives the DVA price. The three DVA quantiles are at confidence levels 0.68, 0.95 and 0.997 (from bottom to top).

\item[Q:] An this second figure you are bringing up now?

\item[A:] This one [showing Figure \ref{fig:FVA_qs} on the tablet screen] is the same example as in the earlier figure but now we are plotting the 
funding price component (price without funding minus price with funding)  for a 10y receiver IRS cleared by a CCP with initial margins at different levels of safety. As before level of safety is meant as a confidence level for the VaR that is used to establish the initial margin. We could use expected shortfall (ES) as well. 

\item[Q:] I see. It makes sense that funding costs increase with the level of safety of initial margins. Safer initial margins are larger, corresponding to larger VaR or ES due to larger confidence levels. I don't need to see the exact maths to see that the pattern makes sense.  What next?

\begin{figure}
\begin{center}
\includegraphics[scale=1]{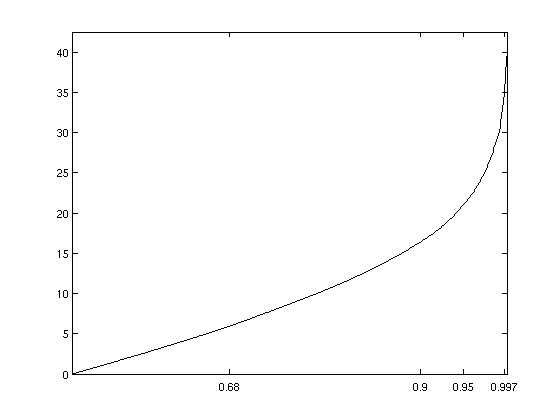} 
\end{center}
\caption{Funding price component (price with funding minus price without funding)  for a 10y receiver IRS cleared by a CCP with initial margins at different levels of safety}\label{fig:FVA_qs}
\end{figure}

\newpage

\item[A:] I am going to show you what happens if we remove collateralization  (variation and initial margins) entirely. Look at this [brings Figure \ref{fig:WWR_cdfva} on screen]. Here there is no margin and we have CVA, DVA and funding cost contributions. 
In particular, before you ask, on the x-axis you see different values for the instantaneous correlation between short term interest rates driving the IRS and default intensities. The black continuous line is the IRS price inclusive of CVA and DVA but without funding costs, the two components being separately plotted in black dotted lines (CVA adjusted price is the lower line, and DVA adjusted the upper line). The red continuous line is the price corrected also for funding costs. 

\item[Q:]
It looks like the funding wrong way risk is less relevant than the credit one?

\item[A:]
This may have to do with the funding being a diluted effect whereas the credit event may be a strong discontinuity that may be amplified more dramatically by WWR. However in reality the situation is more complex and you should really have a look at
\cite{CCPPallaBr2013}

\begin{figure}
\begin{center}
\includegraphics[scale=1]{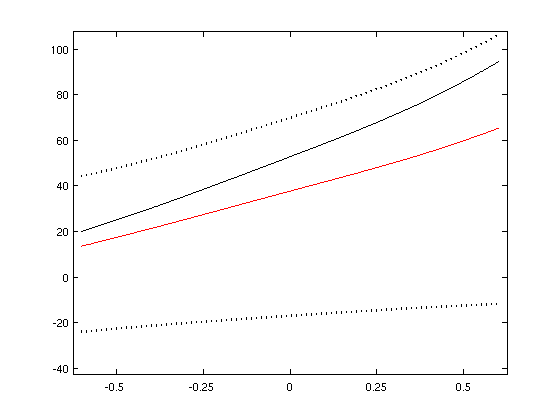} 
\end{center}
\caption{Funding price component (price with funding minus price without funding)  for a 10y receiver IRS cleared by a CCP with initial margins at different levels of wrong way risk correlation}\label{fig:WWR_cdfva}
\end{figure}

\newpage

\item[Q:] Let me be the judge of that. Let's complete the numerical example.

\item[A:] All right. Next plot coming up [shows Figure \ref{fig:CDFVA_alfa}]. Here I am showing you a case without initial margin and with variation margins covering different fractions of the mark to market (the fraction is in the x axis). So if the fraction is 0 there is no variation margin, if it is 1 the variation margin covers the whole mark to market, all other cases are in-between. 

\item[Q:] What are the different lines? 

\item[A:] 
The black continuous line is the IRS price inclusive of CVA and DVA but without funding costs, the two adjusted components are separately plotted in black dotted lines (CVA is the lower line, and DVA the upper line). The red continuous line is the price corrected also for funding costs.

\item[Q:] This makes sense too. As you cover more and more of the mark to market with the variation margin, moving from left to right, all lines converge to 0, meaning that residual CVA DVA and funding become very small and approach zero  with perfect variation margin and zero initial margin. So here we get a feeling for what happens when initial margin is neglected and variation margin may vary in effectiveness?

\begin{figure}
\begin{center}
\includegraphics[scale=1]{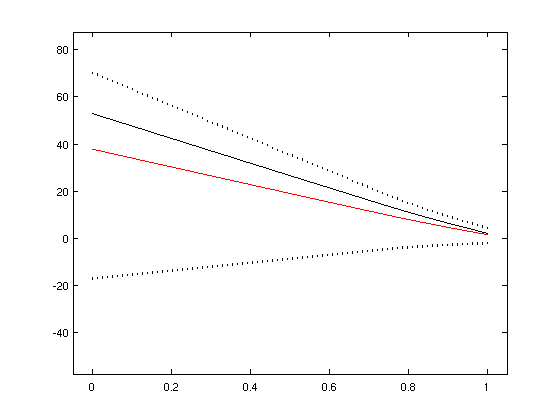} 
\end{center}
\caption{CVA, DVA and Funding inclusive price (y axis) under variation margin at different levels of mark to market coverage (x axis). }\label{fig:CDFVA_alfa}
\end{figure}

\newpage

\item[A:] Precisely. Finally I am going to show you one more figure: [brings on the screen Figure \ref{fig:CDFVA_qs}]

\item[Q:] Are you getting the initial margin back into the picture now?

\item[A:] Correct. Now we have full variation margin (fraction 1 of mark to market) for different confidence levels of the initial margins. The black continuous line is the IRS price inclusive of CVA and DVA but without funding costs, the two adjusted components are separately plotted in black dotted lines (CVA is the lower line, and DVA the upper line as before). The red continuous line is the price corrected also for funding costs. I haven't mentioned that "mark to market" should be a little more qualified, but you find all information in the full paper I mentioned earlier. 

\item[Q:] Any more comments on the background of this particular example?

\item[A:] Well in the examples I took the funding rates to be asymmetric and somehow extreme: the borrowing rate spread is equal to the credit spread of "I", the lending one to $0$. Overall from the first figure \ref{fig:DVA_qs} we have seen that to get rid of rare events on DVA on needs to boost the initial margin, resulting in relevant funding costs. Remember we take a margin period of risk of 5 days. 

\item[Q:] Sounds reasonable, but the good thing is you can get a quantitative assessment here, not just qualitative.

\item[A:] In the second Figure \ref{fig:WWR_cdfva} we saw that somehow WWR is less relevant for funding than for credit, with the caveat that these risks may be interconnected so this is to be qualified in a sense made precise again in the full paper. 

\item[Q:]  Again reasonable, as we already commented earlier. 

\item[A:] Then we analyzed in the remaining figures what happens when moving from no collateral at all to just variation margins at different levels of coverage (without initial margins), and then at the complete picture with full variation margins but different levels of safety on initial margins. We saw the impact of funding to be relevant, the more so if one keeps in mind that funding is a diluted effect and one would ususally expect it to be limited. 

\item[Q:] I can see that the analytics and the kind of approach we discussed allow for a quite precise analysis of the credit, collateral, funding and margining patterns. Interesting. 

\item[A:] The paper \cite{CCPPallaBr2013} also mentions MVA, Margin Valuation Adjustement or something.

\item[Q:] I think we already have too many *VA's and some are not separable, as we discussed, so isn't the introduction of further VA's dangerous?

\item[A:]  They have to be handled carefully and with proper analysis, that's for sure.

\begin{figure}
\begin{center}
\includegraphics[scale=1]{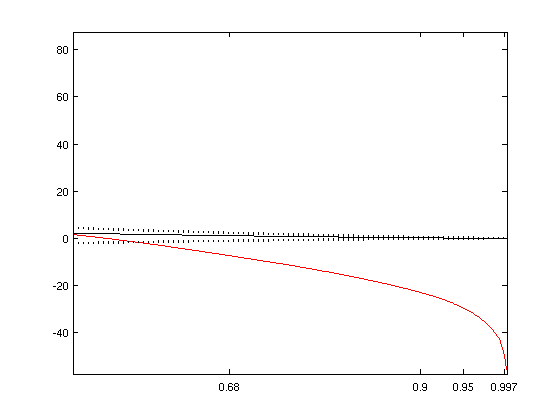} 
\end{center}
\caption{CVA, DVA and Funding inclusive price (y axis) under full variation margin and different confidence levels for the initial margin  (x axis). }\label{fig:CDFVA_qs}
\end{figure}

\item[Q:] So we are back there then... ok this looks like a good place to stop.

\item[A:] It's fine but let me say we haven't discussed multiple curves. 

\item[Q:] Right... Tomorrow?

\item[A:] Can we do that next week please? I need a break!

\item[Q:] Of course, and Thank You!

\end{itemize}

\section{Multiple Discount Curves and Credit-Collateral-Funding}

\begin{itemize}

\item[Q:] [Three days later] Multiple curves: How do they connect with all you told me so far?

\item[A:]  The connection is made quite explicit in the paper \cite{BrigoPalla13}.

\item[Q:] What's the story?

\item[A:] The "story" is that the classical relationship between forward LIBOR rates and zero coupon bonds is not working anymore. In particular, interpreting the zero-coupon curves as risk-free rate curves is no longer tenable. Likewise, interpreting the LIBOR rates as the simply compounded rates that underlie risk-free one-period swaps does not work anymore. To include risks that are now heavily affecting interest rates, one needs to design a theory that includes explicit cash flows accounting for default closeout and also for costs of funding for the hedging portfolio, and costs of funding for collateral (margining).

\item[Q:] Even leaving aside the LIBOR rigging scandal... what are LIBOR rates these days? I'm not sure anymore...

\item[A:] LIBOR is now a rate assigned by the market and not derived by risk-free forward rate agreements or risk-free one-period swaps.

\item[Q:] How does this connect to our discussion?

\item[A:] By including credit, collateral and funding effects in valuation one obtains  the master equation seen earlier, for example, in \cite{BrigoPallaPerini,BrigoPallaPerini2}. In the specific case where collateral is a fraction of current all-inclusive mark-to-market, one obtains a simpler pricing equation based on a hedge-funding-fees equivalent measure and on a generalized dividend process inclusive of default risk, treasury funding and collateral costs. 

\item[Q:] hedge funding fees equivalent measure???

\item[A:] Remember our discussion on Q vs P [in \cite{Brigo2011faq}]? Now we have new Q's where assets evolve with local returns given by funding rates for the hedging portfolios...

\item[Q:] Ah...

\item[A:] Again in \cite{BrigoPalla13} they apply these approximations to the money market, in order to evaluate collateralized interest-rate derivatives.

They look at defining new building blocks that will replace the old unobservables-based ones (such as risk free rate zero coupon bonds, risk-free one period swap rates, etc). Such new instruments are based on the collateral rate, which is an observable rate, since it is contractually defined by the CSA as the rate to be used in the margining procedure. 

\item[Q:] ISDA and CCPs are helping there...

\item[A:]
The collateralized zero-coupon bond can be proxied by single-period Overnight Indexed Swaps (or by quantities bootstrapped from multi-period OIS) when we accept to approximate a daily compounded rate with continuously compounded rates. They then define fair OIS rates at inception in terms of collateralized zero coupon bonds.

\item[Q:] And so LIBOR becomes...

\item[A:] This setup allows us to define new forward LIBOR rates as equilibrium rates in a collateralized one-period swaps.
The resulting rates depend on the collateralized coupon-bearing bonds. They  also define a collateral based forward measure where forward LIBOR rates are expected values of future realized LIBOR rates. 

\item[Q:] Quite a market--based model...

\item[A:] They then hint indeed at the development of a market model theory for the new  collateral-inclusive forward LIBOR rates and to a forward rates theory for the OIS-based instantaneous forward rates.

\item[Q:]  And this is where the multiple curve picture finally shows up I guess

\item[A:] Indeed, they have a curve with LIBOR based forward rates, that are collateral adjusted expectation of LIBOR market rates that are taken as primitive rates from the market, and they have instantaneous forward rates that are OIS based rates.

\item[Q:] Quite reasonable actually. 

\item[A:] OIS rates are driven by collateral fees, whereas forward LIBOR rates are driven both by collateral rates and by the primitive LIBOR market rates. 

\item[Q:] Ok this is a general framework and is very interesting, but are there specific models discussed?

\item[A:] They approach this by introducing a dynamical multiple-curve model for OIS and LIBOR rates, by reformulating the parsimonious HJM model by \cite{Moreni2010,Moreni2012} under the new pricing framework.

\item[Q:] So those are the models they enrich with the credit, collateral and funding analysis?

\item[A:] Correct. They focus on uncollateralized, partially-collateralized and over-collateralized contracts. With partial collateralization they evaluate the adjustment needed by pricing equations to include the corrections coming from Treasury cash and hedging funding.

In particular, forward LIBOR rates associated to partially collateralized one-period swap contracts acquire a covariance term that can be interpreted as a convexity adjustment.

\item[Q:]  What happens with CCPs?

\item[A:] In that paper CCPs initial margins are modeled too. This leads to a generalization of the master formula.
However,  a new difficulty arises. If CCP cleared contracts will take over most of the market, then most of  the hedging activity will be implemented in terms of such products, bringing in funding costs evaluation at the level of the probability measure definition, since the very instruments used to hedge will embed a funding cost.  From a practical point of view, this means  that we may need to adjust OIS curves for funding costs when bootstrapping the discount curves (think of the Futures basis as a similar example).

The paper also discusses credit spreads. It is argued that credit spreads should be calibrated via Credit Default Swaps or Defaultable Bonds. However, again, this should be a global calibration because CDS are collateralized and are in principle priced with the same general formula, inclusive of collateral and funding, that is used for all other deals.

\item[Q:]  Another "Global Calibration"???

\item[A:] Well, bonds are not collateralized and are funded more heavily, so funding risk is also there. This global CDS or Bond calibration is not done usually even though interpreting CDS or Bonds as sources of pure credit risk calibration may lead to important errors. 

\item[Q:] What about funding more specifically?

\item[A:] The term structure of funding rates depends on the funding policy and is model dependent. Stripping it directly from market liquid instruments is very difficult, especially because the interbank market is no longer (fully) representative of such costs. Collateral portfolios play now a key role too. Default intensities, collateral rates and liquidity bases will be key drivers of the funding spreads. Finally, one may  consider using Value-at-Risk measures to determine the effective fraction of mark-to-market one should hold as collateral or even the initial margin, as in our examples above, and introduce collateral haircuts.  

\item[Q:] Quite a composite picture!

\item[A:] At least their final interest-rate curves are consistently explained by such effects and based on market observables. 

\item[Q:] I'll concede this seems a paper worth reading as it seems to put all the pieces together really.

\item[A:] Well I wouldn't say "all", but indeed a lot... and this looks like a good place for a break. [Smiling but looking tired]

\item[Q:] Absolutely. As usual I owe you, big time. May I invite you for dinner? 

\end{itemize}


\newpage

\begin{thebibliography}{99}

\bibitem{albanesebrigooertel} Albanese, C., Brigo, D., and Oertel, F. (2013). Restructuring Counterparty Credit Risk. 
International Journal of Theoretical and Applied Finance, Vol. 16, No. 02.

\bibitem{albanesepietronero} Albanese, C., Pietronero, G., and S. White (2011). Optimal Funding Strategies for Counterparty Credit Risk Liabilities. Working paper available at
\emph{http://www.level3finance.com}.




\bibitem{bergman85}
 Bergman, Y. Z. (1995). Option Pricing with Differential Interest Rates.
The Review of Financial Studies, 
Vol. 8, No. 2 (Summer, 1995), pp. 475-500.


\bibitem{bianchetti10} Bianchetti, M. (2010). Two Curves, One Price. Risk, August 2010. 

\bibitem{bianchetti12} Bianchetti, M. (2012). The Zeeman Effect in Finance: Libor Spectroscopy and Basis Risk Management. Available at arXiv.com

\bibitem{BielCrep}
Bielecki, T., and Crepey, S. (2010). Dynamic Hedging of Counterparty Exposure. Preprint.


\bibitem{BielRut} Bielecki, T., Rutkowski, M. (2001).
\emph{Credit risk: Modeling, Valuation and Hedging}. Springer
Verlag .




\bibitem{oecd} Blundell-Wignall, A., and P. Atkinson (2010). Thinking beyond Basel III.
Necessary Solutions for Capital and Liquidity.
OECD Journal: Financial Market Trends, No. 2, Volume 2010, Issue 1,
Pages 9--33.

Available at {\tt http://www.oecd.org/dataoecd/42/58/45314422.pdf}



\bibitem{Brigo2005}
Brigo, D. (2005).  Market Models for CDS Options and Callable Floaters. Risk Magazine, January issue

\bibitem{Brigo2006}
D. Brigo, Constant Maturity CDS valuation with market models (2006). Risk Magazine, June issue. Earlier extended version available at {\tt http://ssrn.com/abstract=639022}

\bibitem{Brigo2011faq} Brigo, D. (2011). Counterparty Risk, Collateral and Funding FAQ,   \\ {\tt{http://ssrn.com/abstract=1955204}}
or {\tt{arxiv.org/abs/1111.1331}}


\bibitem{alfonsi}
Brigo, D., and Alfonsi, A. (2005) Credit Default Swaps Calibration
and Derivatives Pricing with the SSRD Stochastic Intensity Model,
Finance and Stochastic, Vol. 9, N.~1.

\bibitem{BrigoChourBakkar}
Brigo, D., and Bakkar I. (2009).
Accurate counterparty risk valuation for energy-commodities swaps.
{\em Energy Risk}, March issue.

\bibitem{BrigoBuescuMorini}
Brigo, D., and Buescu, C., and Morini, M. (2012).
Counterparty Risk Pricing: Impact of closeout
and first-to-default time. International Journal of Theoretical and Applied Finance,
Vol. 15, No. 06

\bibitem{brigoliu2012} Brigo, D., Buescu, C., Pallavicini, A., and Qing Liu (2012). Illustrating a problem in the self-financing condition in two 2010-2011 papers on funding, collateral and discounting. Available at SSRN.com and arXiv.org

\bibitem{BrigoCapponi}
Brigo, D., and Capponi, A. (2008).
Bilateral counterparty risk valuation with stochastic dynamical models and application to CDSs.
Working paper available at {\tt http://arxiv.org/abs/0812.3705}$\,$.
Short updated version in {\em Risk}, March 2010 issue.

\bibitem{BrigoCapponiPalla}
Brigo, D., Capponi, A., and Pallavicini, A. (2011).
Arbitrage-free bilateral counterparty risk valuation under collateralization and application to Credit Default Swaps.
To appear in {\em Mathematical Finance}.

\bibitem{brigoetalcollateral}
Brigo, D., Capponi, A., Pallavicini, A., and Papatheodorou, V. (2013).
Pricing Counterparty Risk Including Collateralization, Netting Rules, Re-hypothecation and Wrong Way
Risk. International Journal of Theoretical and Applied Finance Vol. 16, No. 02. Working paper available at {\tt http://arxiv.org/abs/1101.3926}


\bibitem{Brigo08}
Brigo, D., and Chourdakis, K. (2008).
Counterparty Risk for Credit Default Swaps: Impact of spread volatility and default correlation.
{\em International Journal of Theoretical and Applied Finance}, {\bf 12}, 7.


\bibitem{BrigoCousot}
D. Brigo, L. Cousot (2006). A Comparison between the SSRD Model and the Market Model for CDS Options Pricing.  International Journal of Theoretical and Applied Finance, Vol 9, n. 3

\bibitem{brigodurand} Brigo, D., and Durand, C. (2014). An actuarial approach to funding costs valuation. In preparation. 

\bibitem{elbachir}
Brigo, D., and  El--Bachir, N. (2010). An exact formula for default swaptions pricing in the SSRJD stochastic intensity model. Mathematical Finance, Volume 20, Issue 3, Pages 365-382.

\bibitem{BrigoMas}
Brigo, D., and Masetti, M. (2005).
Risk Neutral Pricing of Counterparty Risk.
In Counterparty Credit Risk Modelling: Risk Management, Pricing and Regulation,
{\em Risk Books}, Pykhtin, M. editor, London.

\bibitem{BrMer} Brigo, D., Mercurio, F. (2001).
Interest Rate Models: Theory and Practice with Smile,
Inflation and Credit, Second Edition 2006, Springer Verlag, Heidelberg.

\bibitem{BrigoMorini2006}
Brigo, D., Morini, M. (2006)
Structural credit calibration, {\em Risk}, April issue.

\bibitem{BrigoMorini2010}
Brigo, D., and Morini, M. (2010).
Dangers of Bilateral Counterparty Risk: the fundamental impact of closeout conventions. Preprint available at
{\tt ssrn.com} or at {\tt arxiv.org}.


\bibitem{BrigoMorini2010Flux}
Brigo, D., and Morini, M. (2010). Rethinking Counterparty Default, Credit flux, Vol 114,
pages 18--19

\bibitem{BrigoMorini2011Risk}
Brigo, D., and Morini, M. (2011). Close out convention tensions. Risk Magazine, Dec 2011 issue, to appear. 

\bibitem{BMP2013} Brigo, D., Morini, M., and Pallavicini, A. (2013). Counterparty Credit Risk, Collateral and Funding, with Pricing Cases for all Asset Classes. Wiley, Chichester. 


\bibitem{BrigoMoriniTarenghi}
Brigo D., Morini M., and Tarenghi M. (2011).
Equity Return Swap valuation under Counterparty Risk. In: Bielecki, Brigo and Patras (Editors), Credit Risk Frontiers: Sub- prime crisis, Pricing and Hedging, CVA, MBS, Ratings and Liquidity, Wiley, pp 457--484

\bibitem{BrigoPalla07} Brigo, D., and Pallavicini, A. (2007). Counterparty Risk under Correlation between Default and Interest Rates. In: Miller, J., Edelman, D., and Appleby, J. (Editors), Numercial Methods for Finance, Chapman Hall.

\bibitem{BrigoPalla08}
D. Brigo, A. Pallavicini (2008). Counterparty Risk and Contingent CDS under correlation, Risk Magazine, February issue.

\bibitem{BrigoPalla13}
D. Brigo, A. Pallavicini (2013). Interest-Rate Modelling in Collateralized Markets: Multiple curves, credit-liquidity effects, CCPs. SSRN.com and arXiv.org


\bibitem{BrigoPallaPapa}
Brigo, D., Pallavicini, A., and Papatheodorou, V. (2011).
Arbitrage-free valuation of bilateral counterparty risk for interest-rate products: impact of volatilities and correlations, International Journal of Theoretical and Applied Finance, 14 (6), pp
773--802


\bibitem{brigopallatorre} Brigo, D., Pallavicini, A., and Torresetti, R. (2010).
Credit Models and the Crisis: A journey into CDOs, Copulas, Correlations and
Dynamic Models. Wiley, Chichester.

\bibitem{BrigoTarenghi2004}
Brigo, D. and Tarenghi, M. (2004). Credit Default Swap Calibration
and Equity Swap Valuation under Counterparty risk with a Tractable
Structural Model. Working Paper, available at {\tt
www.damianobrigo.it/cdsstructural.pdf}. Reduced version in {\em
Proceedings of the FEA 2004 Conference at MIT, Cambridge,
Massachusetts, November 8-10} and in {\em Proceedings of the
Counterparty Credit Risk 2005 C.R.E.D.I.T. conference}, Venice,
Sept 22-23, Vol 1.

\bibitem{BrigoTarenghi2005}
Brigo, D. and Tarenghi, M. (2005). Credit Default Swap Calibration
and Counterparty Risk Valuation with a Scenario based First
Passage Model. Working Paper, available at {\tt
www.damianobrigo.it/cdsscenario1p.pdf} \ \ Also in: {\em
Proceedings of the Counterparty Credit Risk 2005 C.R.E.D.I.T.
conference}, Venice, Sept 22-23, Vol 1.


\bibitem{burgard} Burgard, C., and Kjaer, M. (2010).
PDE Representations of Options with Bilateral Counterparty Risk and Funding Costs. Available at ssrn.com

\bibitem{Burgard2011} Burgard, C., and Kjaer, M. (2010). Partial
Differential Equation Representations of Derivatives with Counterparty Risk
and Funding Costs. The Journal of Credit Risk. 7 (3) 1--19

\bibitem{BurgardRisk2011} Burgard, C., and Kjaer, M. (2011). In the Balance.
Risk Magazine. October issue.

\bibitem{CannDuff}
Canabarro, E., and Duffie, D. (2004).
Measuring and Marking Counterparty Risk. In Proceedings of the Counterparty Credit Risk 2005 C.R.E.D.I.T. conference, Venice, Sept 22--23, Vol 1.

\bibitem{cannabaro05} Canabarro, E., Picoult, E., and Wilde, T. (2005). Counterparty Risk. {\em Energy Risk}, May issue.

\bibitem{Castagna2011} Castagna, A. (2011). Funding, Liquidity, Credit and Counterparty Risk: Links and Implications,
 Available at http://ssrn.com/abstract=1855028

\bibitem{cesari} G. Cesari, J. Aquilina ,  N. Charpillon, Z. Filipovic, G. Lee and I. Manda (2010).
Modelling, Pricing, and Hedging Counterparty Credit Exposure: A Technical Guide,
Springer Verlag, Heidelberg.

\bibitem{Cherubini2005}
Cherubini, U. (2005). Counterparty Risk in Derivatives and
Collateral Policies: The Replicating Portfolio Approach. In:
\emph{ALM of Financial Institutions} (Editor: Tilman, L.), Institutional Investor Books.

\bibitem{collin} Collin-Dufresne, P., Goldstein, R., and Hugonnier, J. (2002). A general formula for pricing defaultable securities. Econometrica 72: 1377-1407.

\bibitem{Rama2013} Cont, R. and Kokholm, T. (2013). Central Clearing of OTC Derivatives: bilateral vs multilateral netting,
ssrn.com


\bibitem{Crepey2011}
Cr\'epey, S. (2011). A BSDE approach to counterparty risk under funding constraints.
Available at

{\tt grozny.maths.univ-evry.fr/pages$\_$perso/crepey}

\bibitem{Crepey2012a} Cr\'epey, S. (2012). Bilateral  Counterparty Risk under Funding Constraints – Part I: Pricing. Forthcoming in Mathematical Finance

\bibitem{Crepey2012b} Cr\'epey, S. (2012). Bilateral  Counterparty Risk under Funding Constraints – Part II: CVA. Forthcoming in Mathematical Finance


\bibitem{crepey2012c} Cr\'epey S., Grbac, Z., Ngor, N. (2012). 
A multiple-curve HJM model of interbank risk.
{\it Mathematics and Financial Economics} 6(3) 155-190.

\bibitem{crouhy}
Crouhy M., Galai, D., and  Mark, R. (2000). A comparative analysis
of current credit risk models. {\em Journal of Banking and
Finance} 24, 59-117

\bibitem{danziger} Danziger, J. (2010). Pricing and Hedging Self Counterparty Risk.
Presented at the conference ``Global Derivatives", Paris, May 18, 2010.

\bibitem{DuffieHuang}
Duffie, D., and Huang, M. (1996). Swap Rates and Credit Quality. Journal of Finance 51, 921--950.

\bibitem{DuffieZhu}
Duffie, D., and Zhu H. (2010). Does a Central Clearing Counterparty Reduce Counterparty Risk? Working Paper, Stanford University.

\bibitem{ehlers} Ehlers, P. and  Schoenbucher, P. (2006). The Influence of FX Risk on Credit Spreads, ETH working paper, available at defaultrisk.com

\bibitem{elkaroui} El Karoui, N., Peng, S. and Quenez M. (1997). Backward stochastic differential equations in finance, Mathematical Finance,
vol 7, pp 1--71



\bibitem{fries} Fries, C. (2010). Discounting Revisited: Valuation Under Funding, Counterparty Risk and Collateralization. Available at SSRN.com

\bibitem{Fujii2010}
Fujii, M., Shimada, Y., and Takahashi, A. (2010).
Collateral Posting and Choice of Collateral Currency. Available at {\tt ssrn.com}.

\bibitem{Fujii10} 
Fujii, M., Shimada, Y., Takahashi, A. (2010)\\
On the Term Structure of Interest Rates with Basis Spreads, Collateral and Multiple Currencies.\\
Available at \url{http://ssrn.com/abstract=1556487}.

\bibitem{Gregory} J. Gregory (2009). \textquotedblleft Being two faced
over counterparty credit risk\textquotedblright , \emph{Risk Magazine} 22
(2), pages 86-90.

\bibitem{gregorybook}
J. Gregory (2010).
Counterparty Credit Risk: The New Challenge for Global Financial Markets,
Wiley, Chichester.


\bibitem{henrard2007}  Henrard, M. (2007). The Irony in the Derivatives Discounting. Wilmott Magazine,
July 2007, 92-98.

\bibitem{henrard2009} Henrard, M.  (2009). The Irony in the Derivatives Discounting Part II: The Crisis.
Preprint.





\bibitem{isdavarious} ISDA.
``Credit Support Annex'' (1992), `Guidelines for Collateral Practitioners''(1998), ``Credit Support Protocol'' (2002), ``Close-Out Amount Protocol'' (2009), ``Margin Survey'' (2010), ``Market Review of OTC Derivative Bilateral Collateralization Practices'' (2010).
Available at http://www.isda.org






\bibitem{Jamshidian} Jamshidian, F. (2002).
Valuation of credit default swap and swaptions,
FINANCE AND STOCHASTICS, 8, pp 343--371

\bibitem{Jones} Jones, E. P., Mason, S.P., and Rosenfeld, E.
(1984). Contingent Claims Analysis of Corporate Capital Structure:
An Empirical Investigation. {\em Journal of Finance} 39,  611-625

\bibitem{jorion} Jorion, P. (2007). Value at Risk, 3-d edition, McGraw Hill.

\bibitem{keenan} Keenan, J. (2009). Spotlight on exposure. {\it Risk} October issue.


\bibitem{Kenyon}
Kenyon, C. (2010).
Completing CVA and Liquidity: Firm-Level Positions and Collateralized Trades. Available at {\tt arXiv.org}.

\bibitem{Kiff2009}
Kiff, J. (2009). Mitigating Counterparty Credit Risk in OTC Markets: The Basics,
{\tt shadowbankers.wordpress.com}


\bibitem{Leung} Leung, S.Y., and Kwok, Y. K. (2005). Credit Default Swap Valuation with Counterparty Risk. The Kyoto Economic Review 74 (1), 25--45.

\bibitem{liptonsepp}
Lipton A, Sepp A, Credit value adjustment for credit default swaps via the structural default model, The Journal of Credit Risk, 2009, Vol:5, Pages:123-146


\bibitem{LLH} Lo, C. F., Lee H.C. and Hui, C.H.  (2003).
A Simple Approach for Pricing Barrier Options with Time-Dependent
Parameters. {\em Quantitative Finance} 3, 98-107




\bibitem{mcneil} McNeil, A. J., Frey, R., and P. Embrechts (2005). Quantitative Risk Management:
Concepts, Techniques, and Tools, Princeton university press.


\bibitem{mercurio2009} F. Mercurio (2009). Interest Rates and The Credit Crunch: New Formulas and Market
Models. Bloomberg Portfolio Research Paper No. 2010-01-FRONTIERS. Available at SSRN.com



\bibitem{merton74}
Merton R. (1974) On the Pricing of Corporate Debt: The Risk
Structure of Interest Rates. {\em The Journal of Finance} 29,
449-470.

\bibitem{Miller2011}
Miller, R. S. (2011). Conflicts of interest in derivatives clearing,
Congressional Research Service Report


\bibitem{Moreni2010}
Moreni, N. and Pallavicini, A. (2010). 
Parsimonious HJM Modelling for Multiple Yield-Curve Dynamics. 
Accepted for publication in Quantitative Finance.

\bibitem{Moreni2012}
Moreni, N. and Pallavicini, A. (2012). 
Parsimonious Multi-Curve HJM Modelling with Stochastic Volatility, in: Bianchetti and Morini (Editors), 
Interest Rate Modelling After The Financial Crisis, Risk Books.

\bibitem{morini2009} Morini, M. (2009). Solving the Puzzle in the Interest Rate Market. Available at SSRN.com


\bibitem{morinibook} Morini, M. (2011).
Understanding and Managing Model Risk. Wiley.

\bibitem{BrigoMorini2011}
Morini, M. and Brigo, D. (2011). No-Armageddon Measure for Arbitrage-Free Pricing of Index Options in a Credit Crisis,
Mathematical Finance, 21 (4), pp 573--593


\bibitem{MoriniPrampolini} Morini, M. and Prampolini, A. (2011). Risky Funding: A Unified Framework for Counterparty and Liquidity Charges, Risk Magazine, March 2011 issue.

\bibitem{pallaNested}
Pallavicini, A. (2010).
Modelling Wrong-Way Risk for Interest-Rate Products.
Presented at the conference ``6th Fixed Income Conference", Madrid, 23-24 September 2010.

\bibitem{CCPPallaBr2013} Pallavicini, A., and Brigo, D. (2013). CCP Cleared Contracts and Bilateral Trades under CSA with  Pricing of Interest-Rate Derivatives: funding and wrong-way risk with re-hypothecation, funding and wrong way risk: A unified approach. To appear on SSRN.com and arXiv.org

\bibitem{BrigoPallaPerini}
Pallavicini, A., Perini, D., and Brigo, D. (2011). Funding Valuation Adjustment consistent with CVA and DVA, wrong way risk, collateral, netting and re-hypotecation. Available at SSRN.com and arXiv.org


\bibitem{BrigoPallaPerini2}
Pallavicini, A., Perini, D., and Brigo, D. (2012). Funding, Collateral and Hedging: uncovering the mechanics and the subtleties of funding valuation adjustments. Available at SSRN.com and arXiv.org


\bibitem{ParkerMcGarry}
Parker E., and McGarry A. (2009)
The ISDA Master Agreement and CSA: Close-Out Weaknesses Exposed in the Banking Crisis and Suggestions for Change.
Butterworths Journal of International Banking Law, 1.

\bibitem{pengnonlin} Shige Peng (2010). Nonlinear Expectations and Stochastic Calculus under Uncertainty, available at arXiv.org


\bibitem{Piron2012} Piron, B. (2012). Why collateral and {CCPs} can be bad for your wealth. Sungard Adaptiv White Paper,
2012

\bibitem{Piter2010}
Piterbarg, V., (2010).
Funding beyond discounting: collateral agreements and derivatives pricing. Risk Magazine, February 2010.


\bibitem{Piterbarg2012}
Piterbarg, V. (2012). Cooking with collateral. Risk Magazine, Vol 8.




\bibitem{RutkowskiFunding} Bielecki, T., and Rutkowski, M. (2013). Valuation and hedging of OTC contracts with funding costs, collateralization and counterparty credit risk: Part 1. arXiv.org

\bibitem{Tabb2011} Rhode, W. (2011). 
European Credit and Rates Dealers 2011: Capital,
Clearing and Central Limit Order Books, Tabb Group,
2011.


\bibitem{rosenpyk} Rosen, D., and Pykhtin, M. (2010). Pricing Counterparty Risk at the Trade Level and CVA Allocations. Journal of Credit Risk, vol. 6 (Winter 2010), pp. 3-38.

\bibitem{Singh2011b}
Singh, M. (2011). Making OTC Derivatives Safe - A Fresh Look,
IMF Working Paper, WP/11/66



\bibitem{sorensen} Sorensen, E.H., and Bollier, T. F. (1994).
Pricing Swap Default Risk. {\em Financial Analysts Journal}, 50.
23--33.




\bibitem{TorreJRMFI} Torresetti, R., Brigo, D., and Pallavicini, A (2009). Risk-neutral versus objective loss distribution and CDO tranche valuation. Journal of Risk Management in Financial Institutions.Vol. 2, 2 175--192.

\bibitem{watt} Watt, M. (2011).
Corporates fear CVA charge will make hedging too expensive.
Risk Magazine, October issue.

\bibitem{Wayne2012}
Wayne, H. (2012). Basel 3, Dodd Frank and EMIR,
Citigroup Presentation.


\bibitem{WeeberRobson}
Weeber, P., and Robson E. S. (2009)
Market Practices for Settling Derivatives in Bankruptcy. ABI Journal, 9, 34--35, 76--78.


\end{thebibliography}
\end{document}